\documentclass[useAMS]{mn2e}
\usepackage[dvips]{graphicx}
\newif\ifAMStwofonts

\begin{document}

\title[Mass accretion in a clumpy ISM]{Radiation drag driven mass accretion in clumpy interstellar medium:
implications for the supermassive black hole-to-bulge relation
}
\author[N. Kawakatu and M. Umemura]
       {Nozomu Kawakatu and Masayuki Umemura \\
        Center for Computational Physics, University of Tsukuba, Tsukuba, Ibaraki 305, Japan}
\date{Accepted 2001 September 20.
      Received 2001  August 17;
      in original form 2001 May 30}

\pagerange{\pageref{firstpage}--\pageref{lastpage}}
\pubyear{2001}

\maketitle

\label{firstpage}

\begin{abstract}
We quantitatively scrutinize the effects of the radiation drag 
arising from the radiation fields in a galactic bulge
in order to examine the possibility that the radiation drag
could be an effective mechanism to extract angular momentum
in a spheroidal system like a bulge
and allow plenty of gas to accrete onto the galactic center.
For this purpose, we numerically solve the relativistic radiation 
hydrodynamical equation coupled with the accurate radiative transfer and
quantitatively assess the radiation drag efficiency.
As a result, we find that in an optically thick regime
the radiation drag efficiency is sensitively dependent on the density distributions 
of interstellar medium (ISM).
The efficiency drops according to $\tau_{\rm T}^{-2}$ 
in an optically thick {\it uniform} ISM, 
where $\tau_{\rm T}$ is the total optical depth of the dusty ISM, 
whereas the efficiency remains almost constant at a high level 
if the ISM is {\it clumpy}.
Hence, if the bulge formation begins with a star formation event in a clumpy ISM, 
the radiation drag will effectively work 
to remove the angular momentum and the accreted gas may form
a supermassive black hole. As a natural consequence,
this mechanism reproduces a putative linear relation between 
the mass of a supermassive black hole and the mass of a galactic 
bulge, although further detailed modeling for stellar evolution is
required for the more precise prediction.
\end{abstract}


\newpage
\section{Introduction}
\label{INTRO}
Recently, Kormendy $\&$ Richstone (1995) have pioneeringly 
suggested that the mass of a supermassive black hole (BH)
does correlate linearly with the mass of the hosting bulge.
(It is noted that the term of a bulge is used to mean a whole galaxy 
for an elliptical galaxy in this paper as is  often so.)
Further high-quality observations of the galactic center 
using stellar dynamics, gasdynamics, and maser dynamics 
(Miyoshi et al. 1995; Magorrian et al. 1998; Richstone et al. 1998;
Ho 1999; Wandel 1999; Kormendy \& Ho 2000;
Ferrarese et al. 2000; Gebhardt et al. 2000a;
Sarzi et al. 2001; Merritt \& Ferrarese 2001a) 
allow us to make a detailed demography of supermassive BHs. 
The recent findings are the following:
(1) The BH mass exhibits a linear relation to the bulge mass 
for a wide range of BH mass with a median BH mass fraction
of $f_{\rm BH}\equiv M_{\rm BH}/M_{\rm bulge}=0.001-0.006$
 (Kormendy \& Richstone 1995; Richstone et al. 1998;
Magorrian et al. 1998; Gebhardt et al. 2000a;
Ferrarese \& Merritt 2000; Merritt \& Ferrarese 2001a).
(2) The BH mass correlates with the velocity dispersion of bulge
stars with a power-law relation as $M_{\rm BH} \propto \sigma^n$,
$n=3.75$ (Gebhardt et al. 2000a) or 4.72 (Ferrarese \& Merritt 2000;
Merritt \& Ferrarese 2001a, 2001b).
(3) $f_{\rm BH}$ tends to grow with the age of youngest stars in a bulge
until $10^9$ yr (Merrifield, Forbes \& Terlevich 2000).
(4) In disc galaxies, the mass ratio is significantly smaller than 0.001
if the disc stars are included (Salucci et al. 2000; Sarzi et al. 2001).
(5) For quasars, the $f_{\rm BH}$ is on a similar level to
that for elliptical galaxies (Laor 1998; Mclure \& Dunlop 2001a; Wandel 2001).
(6) The $f_{\rm BH}$ in Seyfert 1 galaxies is under debate, which may be
considerably smaller than 0.001 (Wandel 1999; Gebhardt et al. 2000a) or
similar to that for ellipticals (McLure \& Dunlop 2001a,2001b; Wandel 2001),
while the BH mass-to-velocity dispersion relation
in Seyfert 1 galaxies seems to hold good in a similar way to elliptical
galaxies (Gebhardt et al. 2000b; Nelson 2000; Ferrarese et al. 2001).
These BH-to-bulge correlations suggest 
that the formation of a supermassive BH is physically 
connected with the formation of a galactic bulge.

So far, very little is understood about the physical 
mechanism to produce such correlations, 
although some theoretical models has been proposed (Silk $\&$ Rees 1998; 
Ostriker 2000; Adams, Graff $\&$ Richstone 2001).
Recently, as a possible mechanism to work in a spheroidal system,
Umemura (2001) has considered the effects of radiation drag.
The radiation drag is a relativistic effect, 
which may extract the angular momentum effectively 
in a spheroidal system like a bulge,
so that plenty of interstellar medium (ISM) could accrete on to the galactic center.
Obviously, the radiation drag is inefficient in present-day
elliptical galaxies or galactic bulges, since they possess little ISM.
If the contents of a supermassive BH
are initially in the form of ISM, however, the bulge must have been optically thick 
in the early stage:
$$
\tau \approx  \chi \rho r_{\rm b}= 1.0 
\left({\chi \over 100\, {\rm cm^{2}g^{-1}}}\right)
\left({M_{\rm gas} \over 10^9 M_\odot}\right)
\left({r_{\rm b} \over {\rm 3kpc}}\right)^{-2},
$$
where $\chi$ is the mass extinction coefficient due to dust opacity, 
$\rho$ is the density of the ISM, $M_{\rm gas}$ is the mass of the ISM, 
and $r_{\rm b}$ is the bulge radius.
If a considerable amount of gas is expelled by a galactic wind at some stage,
the optical depth should be still larger before the wind.
If the radiation drag works efficiently in an optically thick medium, 
the rate of mass  accretion induced by the radiation drag is maximally
$L_{\rm bol}/c^2$ (Umemura, Fukue \& Mineshige 1997, 1998; 
Fukue, Umemura \& Mineshige 1997), where $L_{\rm bol}$ is 
the bolometric luminosity.
Umemura (2001) has found that, if the maximal drag efficiency is achieved,
the resultant BH-to-bulge mass ratio is basically determined 
by the energy conversion efficiency of the nuclear fusion from hydrogen to helium, i.e., 0.007.
However, it is not very clear whether this mechanism really works efficiently in realistic situations. 

In this paper, we investigate in detail 
the efficiency of the radiation drag in an optically thick ISM
to test whether the radiation drag model is promising to account for 
the putative BH-to-bulge correlations.
In particular, we concentrate our attention on the effects of
the inhomogeneity in the ISM.
The model for the chemical evolution of elliptical galaxies 
suggests that an elliptical galaxy is initiated by a starburst 
in its early stage of ($< 10^7$yr), and evolves passively after 
a galactic wind event at a few $\times$$10^8$ yr
(Arimoto $\&$ Yoshii, 1986, 1987; Kodama $\&$ Arimoto, 1997; Mori et al. 1997).
Also, in nearby starburst galaxies that have been studied, 
the ISM is highly clumpy (Sanders et al. 1988; Gordon et al. 1997).
Thus, if we consider the radiation drag in the early phase of bulge evolution,
we should consider an inhomogeneous optically thick ISM. 
In this paper, to elucidate the mutual effect between the clumpiness of ISM 
and the optical depth on the radiation drag efficiency,
we build up a simple model of the bulge system and accurately solve 
the radiation transfer in a clumpy ISM.

The paper is organized as follows.
In Section 2, we construct the model of a galactic bulge.
In Section 3, the basic equations for the ISM are provided.
In Section 4,  the angular momentum transfer efficiency is assessed 
in a {\it uniform} ISM.
In Section 5, we investigate 
the angular momentum transfer efficiency by solving the radiation 
transfer in a {\it clumpy} ISM, and elucidate the relationship 
between the clumpiness
of the ISM and the angular momentum transfer efficiency.  
In Section 6, we give implications for the correlation 
between the supermassive BH  mass and the bulge mass.
In addition, we discuss further effects that would give 
significant influence on the BH mass.
Section 7 is devoted to our conclusions.

\section{Model}
\label{MD}
We assume that a spherical galactic bulge consists of three components, that is, 
dark matter, stars, and a dusty ISM. 
The bulge radius $r_{\rm b}$ is set to be 1-10kpc by taking account of 
the observed sizes of elliptical galaxies or bulges in spiral galaxies.
Dark matter is distributed uniformly inside the bulge.
The mass of dark matter component, $M_{DM}$, within the bulge size 
is equal to the stellar mass of the galactic bulge, $M_{\rm bulge}$. 
As for the stellar component, we assume star clusters with 
a specific stellar initial mass function (see below).
The star clusters are distributed uniformly inside the galactic bulge.
Hereafter, `a star' in this paper means `a star cluster'.
Here, $N_{*} (=100)$ stars are distributed randomly.
For a dusty ISM, we consider the two cases:
one is a uniformly distributed ISM, and the other is a clumpy ISM.
In the case of clumpy ISM, $N_{\rm c}(=10^{4})$ identical clouds are 
distributed randomly.
The density $\rho_{\rm gas}$ in a cloud is assumed to be uniform.
The size of a gas cloud, $r_{\rm c}$, is a parameter.
Then,the optical depth of a gas cloud is $\bar{\tau}=\chi \rho_{\rm gas}r_{\rm c}$,
where $\chi$ is the mass extinction.
We suppose that the stars and the ISM clouds corotate with the angular velocity 
corresponding to the angular momentum obtained by the tidal torque at the linear 
stage of density fluctuations.
Quantitatively, the angular momentum is given 
by the spin parameter $\lambda=(J_{\rm T}|E_{\rm T}|^{1/2})
/(GM_{\rm T}^{5/2})=0.05$, where 
$J_{\rm T}$, $E_{\rm T}$, and $M_{\rm T}$ are respectively 
the total angular momentum, energy, and mass
(Barnes \& Efstathiou 1987; Heavens \& Peacock 1988).
Here, the rigid rotation is assumed.

The mass range of galactic bulges is postulated to be $10^{6-13}M_{\odot}$. 
(However, in the present analysis, it is not very 
important to specify $M_{\rm bulge}$, 
because the results are scaled with $M_{\rm bulge}$ as shown below.)
The total mass of the ISM, $M_{\rm gas}$, and the mass of each gas cloud, $m_{\rm c}$,
are parameters.
If dust opacity as well as Thomson scattering is considered,
the mass extinction is expressed by
$\chi = (n_{\rm e}\sigma_{\rm T}+n_{\rm d}\sigma_{\rm d})/
(\rho_{\rm g}+\rho_{\rm d})$,
where $\sigma_{\rm T}$ is the Thomson scattering cross section, 
$n_{\rm e}$ is the electron number density, $\rho_{\rm g}$ is the 
gas density, and
$n_{\rm d}$, $\sigma_{\rm d}$, and $\rho_{\rm d}$ are respectively
the number density, cross-section, and mass density of dust grains.
If we take the dust-to-gas mass ratio $ f_{\rm dg}$ of $\sim 10^{-2}$, 
then the opacity ratio is 
$$
\frac{n_{\rm d}\sigma_{\rm d}}{n_{\rm e}\sigma_{\rm T}}
=1.9\times 10^{2}\left(\frac{a_{\rm d}}{\rm 1\mu m} \right)^{-1}
\left(\frac{\rho_{\rm s}}{\rm g~cm^{-3}}\right)^{-1}
\left(\frac{f_{\rm dg}}{10^{-2}} \right),
$$
where $a_{\rm d}$ is the grain radius and $\rho_{\rm s}$ is the 
density of solid material within the grain.
Thus, we find $n_{\rm d}\sigma_{\rm d}\gg n_{\rm e}\sigma_{\rm T}$
in the situations of interest. 
Hence, in this paper, we evaluate the mass extinction by
$\chi=n_{\rm d}\sigma_{\rm d}/\rho_{\rm g}$.

Finally, as for the stellar evolution, we assume a Salpeter-type initial mass function
 (IMF) as $\phi =A(m_{*}/M_{\odot})^{-1.35}$ 
for a mass range of $[m_l, m_u]$.
In the present analysis, we consider an initial starburst and 
the subsequent passive evolution of stars. 
The upper mass limit is inferred to be around $40M_{\odot}$ in starburst regions (Doyon, Puxley \& 
Joseph 1992).
As for the lower mass limit, some authors suggest that
the IMF in starburst regions is deficient in low-mass stars, 
with a cut-off of about $2-3M_{\odot}$ (Doane \& Mathews 1993; 
Charlot et al. 1993; Hill et al. 1994).
In this paper, we assume $m_l=2M_\odot$ and $m_u=40M_\odot$ 
in the early stage of bulge formation. However,
it should be also kept in mind that the lower mass limit
is under debate; the claimed lower cut-off
could be due to the magnitude limit effect (Selman et al. 1999),
and also recently sub-solar-mass stars are found
in a starburst region in our galaxy, NGC 3603 (Brandl et al. 1999).
In order to incorporate the stellar evolution, we adopt the mass-luminosity relation $(\ell_{*}/L_{\odot})=(m_{*}/M_{\odot})^{3.7}$,
and the mass-age relation $\tau_{*}=1.1\times 10^{10}(m_{*}/M_{\odot})^{-2.7}{\rm yr}$ 
(Lang 1974), where $m_{*}$, $\ell_{*}$, and $\tau_{*}$ are respectively 
the stellar mass, luminosity, and age. 

\section{Basic equations}
\label{BES}
As a relativistic result of radiative absorption and 
subsequent re-emission, the radiation fields exert a drag force on 
moving material in resistance to its velocity.
This radiation drag extracts angular momentum from the ISM, 
thereby allowing the gas to accrete on to the galactic center.
The radiation drag is an effect of $O(v/c)$, but it could 
provide a key mechanism for angular momentum transfer in intense radiation fields.
We put the origin at the center of the bulge, and adopt cylindrical 
coordinate $r$, $\phi$, and $z$, where $z$-axis is the rotation axis of stars and gas.
The components of the specific radiation force that is exerted 
on moving fluid elements with velocity $\it{\mathbf{v}}$ are given by
\begin{equation}
f_{\rm r}=\frac{\chi}{c}(F^{\rm r}-v_{\rm r}
E-v_{\rm r}P^{\rm rr}-v_{\rm \phi}P^{\rm \phi r}-v_{\rm z}P^{\rm zr}),
\end{equation}
\begin{equation}
f_{\rm \phi}=\frac{\chi}{c}(F^{\rm \phi}-v_{\rm \phi}
E-v_{\rm \phi}P^{\rm \phi \phi}-v_{\rm r}P^{\rm r \phi}-v_{\rm z}P^{\rm z\phi}),
\end{equation} 
and
\begin{equation}
f_{\rm z}=\frac{\chi}{c}(F^{\rm z}-v_{\rm z}E-v_{\rm z}
P^{\rm zz}-v_{\rm r}P^{\rm rz}-v_{\rm \phi}P^{\rm \phi z})
\end{equation}
(Mihalas $\&$ Mihalas 1984) in the $r$-, $\phi$-, and $z$- directions respectively.
Here, $E$ is the radiation energy density, $F^{\alpha}$ 
is the radiation flux, and $P^{\alpha \beta}$ is the radiation stress tensor
where the non-diagonal components are null owing to the present symmetry.

Using equations (1)$-$(3), we have the radiation hydrodynamical equations to $O(v/c)$ as
\begin{equation}
\frac{dv_{\rm r}}{dt}=\frac{v_{\rm \phi}^{2}}{r}-f_{\rm g}^{\rm r}+\frac{\chi}{c}
[F^{\rm r}-(E+P^{\rm rr})v_{\rm r}]
\end{equation}
\begin{equation}
\frac{1}{r}\frac{d(rv_{\rm \phi})}{dt}=\frac{\chi}{c}[F^{\rm \phi}-
(E+P^{\rm \phi \phi})v_{\rm \phi}]
\end{equation}   
\begin{equation}
\frac{dv_{\rm z}}{dt}=-f_{\rm g}^{\rm z}+\frac{\chi}{c}[F^{\rm z}-(E+P^{\rm zz})v_{\rm z}]
\end{equation}
where $f_{\rm g}^{\rm r}$ and $f_{\rm g}^{\rm z}$ are respectively $r$- and $z$- component
of the gravitational force.

The azimuthal equation of motion (5) is the equation of angular momentum transfer.
This equation implies that the radiation flux force 
(the first term on the right-hand side) makes fluid elements 
tend to corotate with stars, whereas the radiation drag 
(the second term on the right-side) works to extract the 
angular momentum from gas.
Therefore, the gain and loss of total angular momentum is determined 
by equation (5).

\section{Uniform ISM}
\label{UI}
In this section, we consider the angular momentum transfer by radiation drag in a 
uniform ISM.
First, we analytically 
calculate the radiation fields produced by spherically distributed stars 
in an optically thin regime, and assess the angular momentum loss rate 
$\dot{J}$ and the mass accretion rate $\dot{M}$.
Next, we extend the analysis to an optically thick regime.
Then, the relationship between the optical depth 
of a dusty ISM and the angular momentum transfer efficiency is derived.

\subsection{Optically thin regime}
For uniform distributions of stars, the radiation fields inside 
the bulge are analytically integrated in an optically thin regime to be 
\begin{eqnarray}
cE &=&\frac{3L_{\rm bulge}}{2\pi r_{\rm b}^{2}}
\left(1-\frac{\pi}{4}\frac{r}{r_{\rm b}}\right),\\
F^{\rm r}&=& \frac{L_{\rm bulge}}{4\pi r_{\rm b}^{3}}r ,\\
cF^{\rm \phi}&=&\frac{3L_{\rm bulge}}{2\pi r_{\rm b}^{2}}\omega r_{\rm b}
\left(\frac{215}{216}-\frac{49\pi}{192}\right)
\left(\frac{r}{r_{b}}\right)^{2}\nonumber \\
& & {}+\frac{3L_{\rm bulge}}{2\pi r_{\rm b}^{2}}\omega r_{\rm b}\frac{8}{9}
\frac{r}{r_{\rm b}}\left(1-\frac{r}{r_{\rm b}}\right),\\
F^{\rm z} &=& \frac{L_{\rm bulge}}{4\pi r_{\rm b}^{3}}z,\\
cP^{\rm rr}&=&\frac{1}{3}\frac{3L_{\rm bulge}}{2\pi r_{\rm b}^{2}}
\left(1-\frac{3\pi}{16}\frac{r}{r_{\rm b}}\right),\\
cP^{\rm \phi \phi}&=&\frac{1}{3}\frac{3L_{\rm bulge}}{2\pi r_{\rm b}^{2}}
\left(1-\frac{9\pi}{32}\frac{r}{r_{\rm b}}\right),\\
cP^{\rm zz}&=&cP^{\rm \phi \phi},
\end{eqnarray}
where $L_{\rm bulge}$ and $r_{\rm b}$ are the luminosity
and radius of the bulge and $\omega$ is the angular velocity of stars.
These quantities are equivalent to those obtained by 
Fukue, Umemura \& Mineshige (1997), 
except that the radiation flux $F^{\phi}$ in the azimuthal directions is 
generated by the rotation of the bulge.
As seen in equation (5), the flux $F^{\phi}$ works to corotate the ISM with stars 
in contrast to the drag force.
In Fig. 1, we compare both forces exerted on gas per unit mass, 
$f^{\rm drag}=\chi (E+P^{\phi \phi})v_{\phi}/c$ 
and 
$f^{\phi}=\chi F^{\phi}/c$,
where $v_{\phi}=r\omega$.
It is found that $f^{\rm drag}$ overwhelms $f^{\rm \phi}$ everywhere. 
Thus, the optically thin ISM can always lose the angular momentum 
as a result of radiation drag.
The angular momentum loss rate per unit volume per unit time  
$\dot{j}$ is evaluated by
\begin{equation}
\dot{j}=-\frac{L_{\rm bulge}}{c^{2}}\frac{3\chi \omega}{2\pi}xf(x)
\end{equation} 
\begin{figure}
\includegraphics[width=8cm,keepaspectratio]{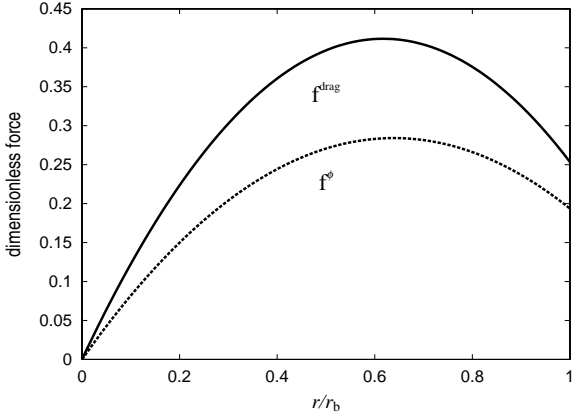}
\caption[fig1.eps]
{The radiation drag force $f^{\rm drag}$ is compared with the flux force $f^{\phi}$ 
in the azimuthal directions.
The $x$-axis is the radial distance normalized by the bulge radius $r_{\rm b}$,
and the radiation force is normalized by 
$(3\chi \omega L_{\rm bulge})/(2\pi c^{2}r_{\rm b})$.
The figure shows that $f^{\rm drag}$ overwhelms $f^{\phi}$ everywhere.
}
\end{figure}
at the point of $x=r/r_{\rm b}$, 
where $f(x)=(\frac{23}{216}+\frac{17\pi}{192})x^{2}+\frac{4}{9}x$.
Equation (14) is integrated over the volume of the bulge to give 
the total angular momentum loss rate $\dot{J}$ as
\begin{equation}
\dot{J}=-\beta \frac{L_{\rm bulge}}{c^{2}}\frac{3\chi \omega}
{2\pi}r_{\rm b}^{3}\rho_{0},      
\end{equation}   
where $\rho_{0}$ is the density of the ISM, and 
$\beta=\int xf(x) dV$ with $dV$ being volume element
in spherical coordinates.
Noting that the initial angular momentum is
\begin{equation}
J_{0}=\frac{2}{5}M_{\rm gas}\omega r_{\rm b}^{2},
\end{equation}
the mass accretion rate is expressed as
\begin{equation}
\dot{M}_{\rm gas}=
-M_{\rm gas}\frac{\dot{J}}{J_{0}}=\eta \frac{L_{\rm bulge}}{c^{2}}\tau_{\rm T}, 
\end{equation}
where $\tau_{\rm T}=\chi \rho_{0}r_{\rm b}=(3\chi M_{\rm gas})/(4\pi r_{\rm b}^{2})$.
The coefficient $\eta$ gives the radiation drag efficiency.
By calculating the constant $\beta$ numerically, $\eta$ is found to be 0.32.

\subsection{Optically thick regime}
In optically thick ISM, the regions where 
the optical depth $\tau _{\rm s}$ from each star is less than unity 
are subject to the radiation drag (Tsuribe $\&$ Umemura 1997).
In this situation, the flux for the region of $\tau_{\rm s}\leq 1$ is given by
\begin{equation}
F^{\rm \phi}=\int ER\omega dV,
\end{equation}
and 
\begin{equation}
F^{\rm drag}=\int (E+P^{\phi \phi})R_{\rm g}\omega dV \simeq \int ER_{\rm g}\omega dV, 
\end{equation}
where $R$ are $R_{\rm g}$ are respectively the distance from the rotational 
axis to a star and to one volume element, 
$dV= r^{2}\sin{\theta}drd\theta d\phi$ in spherical coordinates
for the position of a star.
$E=\ell_{*}/4\pi c r^{2}, P^{\rm \phi \phi} \simeq 0$
with $\ell_{*}$ being the luminosity of a star.
Consequently, the local angular momentum transfer rate by 
equation (5) is
\begin{equation}
\dot{j}=-\frac{\ell_{*}}{c^{2}}\frac{\chi \omega}{6}r_{\rm s}^{3}\rho_{0}.
\end{equation}
This is summed up over all stars to give
\begin{equation}
\dot{J}=-\frac{1}{6}\frac{L_{\rm bulge}}{c^{2}}\chi \omega r_{\rm s}^{3}\rho_{0},
\end{equation}
where $r_{\rm s}$ is the size of the region of $\tau_{s}=1$, 
and $L_{\rm bulge}=N_{*}\ell_{*}$.
Then, the mass accretion rate is given as
\begin{equation} 
\dot{M}_{\rm gas}=-M_{\rm gas}\frac{\dot{J}}{J_{0}}
=\frac{5}{12}\frac{L_{\rm bulge}}{c^{2}}\tau_{\rm T}^{-2}.
\end{equation}
So far, we have considered a rigidly rotating system.
We can also analytically estimate the accretion rate in a system 
with a different rotation law as $v_{\rm \phi}\sim r^{n}$
(e.g. $n=-0.5$ for the Keplarian rotation).
Then, 
the mass accretion rate turns out to be 
\begin{equation} 
\dot{M}_{\rm gas}=-M_{\rm gas}\frac{\dot{J}}{J_{0}}
=\frac{n^{2}(n+4)}{9B}
\left( \frac{\bar{R}_{\rm i}}{r_{\rm b}}\right)^{n-1}
\frac{L_{\rm bulge}}{c^{2}}\tau_{\rm T}^{-2},
\end{equation} 
where $B=\sqrt{\pi}\Gamma(\frac{n+3}{2})/\Gamma(\frac{n+4}{2})$, 
$\bar{R}_{\rm i}$ is the mean distance from the rotational axis to each star, 
and $\Gamma(n)$ is the Gamma function.
It is noted that the difference between (22) and (23) is merely a small factor 
for ordinary rotation laws.
As a result, it is found that the mass accretion rate decreases according 
to $\tau_{\rm T}^{-2}$ in an optically thick regime.
The physical reason is that the larger is the optical depth of the bulge,
the closer to zero is the difference between the velocity of stars 
and the gas velocity, so that the radiation drag efficiency falls.

Combined with the optically thin case, we have the mass accretion rate
in the uniform case as
\begin{equation}
\dot{M}_{\rm gas} \propto \left\{
            \begin{array}{ll}
{\displaystyle \frac{L_{\rm bulge}}{c^{2}}{\tau_{\rm T}}} & (\tau_{\rm T} < 1), \\
{\displaystyle \frac{L_{\rm bulge}}{c^{2}}{\tau_{\rm T}}^{-2}} & (\tau_{\rm T} > 1). 
                 \end{array}   
               \right.
\end{equation}                                 
As readily understood by these results, 
the angular momentum transfer efficiency by the radiation drag is maximum 
when the optical depth of the ISM is around unity.

\section{Clumpy ISM}
\label{ATC}

In the previous section, we considered the angular momentum transfer in a uniform ISM.
In this section, we consider a clumpy ISM.
First, we describe the treatment for the extinction by 
clumpy gas clouds. 
Next, by solving the radiative transfer numerically,
we assess the total angular momentum loss rate $\dot{J}$ and 
the mass accretion rate $\dot{M}_{\rm gas}$.
Then we elucidate the relation between the clumpiness of ISM 
and the angular momentum transfer efficiency quantitatively.

\subsection{Extinction by a clumpy ISM}

In the case of a clumpy ISM, $N_{\rm c}(=10^{4})$ identical clouds 
with optical depth of $\bar{\tau}=\chi \rho_{\rm gas}r_{\rm c}$ are 
distributed randomly.
In an optically thin regime, the radiation fields produced by a star are
\begin{equation}
dE_{0}=\frac{1}{c}\frac{\ell_{*}}{4\pi r^{2}}, 
\,{\it{\mathbf{dF}}}_{0}=\frac{\ell_{*}}{4\pi r^{3}}{\it{\mathbf{r}}}, 
\,dP_{0}^{\rm rr}=dE_{0}, 
\,dP_{0}^{\rm \phi \phi}=dP_{0}^{\rm zz}\simeq 0, 
\end{equation}
where ${\it\mathbf{r}}={\it\mathbf{r}}_{i} - {\it\mathbf{r}}_{j}$ and 
$r = |{\it\mathbf{r}}|$ 
with ${\it\mathbf{r}}_{i}$ being the position of the $i$-th cloud 
and ${\it\mathbf{r}}_{j}$ being the position of $j$-th star.
The physical quantities with suffix $0$ are the radiation fields without extinction.
The corotating flux in the azimuthal direction
and the flux contributing to the radiation drag 
are respectively
\begin{equation}
dF_{0}^{\rm rot}=(dE_{0}+dP_{0}^{\phi \phi})V_{*}, 
\end{equation}
\begin{equation}
dF_{0}^{\rm drag}=-(dE_{0}+dP_{0}^{\rm \phi \phi})v_{\rm gas}, 
\end{equation}
where $V_{*}=r_{*}\omega$ and $v_{\rm gas}=r_{\rm gas}\omega$ 
are the rotational velocities of a star and 
a gas cloud, with $r_{*}$ and $r_{\rm gas}$ being respectively
the distances from the rotational axis to a star and to a gas cloud.
Next, we consider the extinction by dust in clumpy gas clouds.
We calculate the radiation fields by the direct integration
of the radiation transfer (see Fig. 2), 
where the extinction by all intervening clouds from a star to a cloud 
is summed up.
Then, the radiation flux ${\it\mathbf{F}}$, the radiation energy density $E$,
and the radiation stress tensor $P$ are respectively given by
\begin{eqnarray}
{\it{\mathbf{F}}}&=&\sum_{j=1}^{N_{*}} d{\it{\mathbf{F}}}_{0,j}\exp(-\tau_{j}),\\ 
E&=&\sum_{j=1}^{N_{*}} dE_{0,j}\exp(-\tau_{j}),\\
P&=&\sum_{j=1}^{N_{*}} dP_{0,j}\exp(-\tau_{j}), 
\end{eqnarray} 
where $\tau_{j}=2\sum(1-(b/r_{\rm c})^{2})^{1/2}\bar{\tau}$ is 
the optical depth for all intervening clouds 
along the light ray.

\subsection{Angular momentum transfer in a clumpy ISM}
By substituting (26)-(30) for (5),
we can evaluate the total angular momentum loss rate as
\begin{equation}
\dot{J}=\frac{\chi}{c}\sum_{\rm i=1}^{N_{\rm c}}r_{\rm i}
(F^{\rm rot}_{\rm i} - F^{\rm drag}_{\rm i}),
\end{equation}
where $F_{i}^{\alpha}=\sum_{\rm j=1}^{N_{*}}dF^{\alpha}_{0,j}\exp(-\tau _{j})$ 
with $\alpha$ being 'rot' or `drag', and $r_{i}$ is the distance from the rotational axis to 
a gas cloud.
If an optically thick cloud is irradiated by the radiation,
only the optically thin surface layer is subject to radiation 
drag (Tsuribe \& Umemura 1997). Then the surface layer 
is stripped by the drag and simultaneously loses angular momentum.
We assume here that the stripped gas falls on to a central massive object.
Then, with the above angular momentum loss rate, 
we estimate the total mass of the dusty ISM 
assembled on to a central massive object which may evolve into a supermassive BH.
By the equation (31) and the relation $\dot{M}_{\rm gas}/M_{\rm gas}=-\dot{J}/J_{0}$, 
the BH mass $M_{\rm BH}$ is assessed as     
\begin{equation}
M_{\rm BH}=\int_{0}^{t(m_l)}\dot{M}_{\rm gas}dt 
=-\int_{0}^{t(m_l)}M_{\rm gas}\frac{\dot{J}}{J_{0}}dt,
\end{equation}
where $t(m_l)$ is the age of a star with mass $m_l$.

Here, we introduce $\bar{N}_{\rm int}$ as a measure of the clumpiness of ISM.
The $\bar{N}_{\rm int}$ is defined by
\begin{equation}
\bar{N}_{\rm int}=n_{\rm c}\pi r_{\rm c}^{2}r_{\rm b}
=\frac{3}{4}N_{\rm c}\left(\frac{r_{\rm c}}{r_{\rm b}}\right)^{2},
\end{equation}
where $n_{\rm c}=N_{\rm c}/\frac{4}{3}\pi r_{\rm b}^{3}$ is the number density 
of gas clouds.
$\bar{N}_{\rm int}$ means the average number of gas clouds 
that are intersected by a light ray over a bulge radius.
The degree of clumpiness is larger for smaller $\bar{N}_{\rm int}$, 
and $\bar{N}_{\rm int}=\infty$ corresponds to a uniform distribution. 
When $\bar{N}_{\rm int}$ and the optical depth $\bar{\tau}$ of each cloud 
are specified, the total optical depth $\tau_{\rm T}$ is 
given by $\tau_{\rm T}=\bar{N}_{\rm int}\bar{\tau}$.
With changing $\bar{N}_{\rm int}$, we investigate 
the relationship between the clumpiness of ISM and 
the radiation drag efficiency. 

In Fig. 3, for different $\bar{N}_{\rm int}$,
we show the ratio of the resultant BH mass to the bulge mass 
($M_{\rm BH}/M_{\rm bulge}$) against the total optical depth $\tau_{\rm T}$.
In this figure, the dashed line is the analytic solution 
for a uniform ISM corresponding to 
$\bar{N}_{\rm int} = \infty$, i.e. equation (24).
The solid lines show the numerical results for a clumpy ISM,
where the arrows represent $\bar{\tau}=1$ 
for different $\bar{N}_{\rm int}$.
The thick lines show the results for $\bar{N}_{\rm int} \geq 1$
and the thin lines are those for $\bar{N}_{\rm int} < 1$. 
In Fig. 3, several points are clear regarding the effects
of the clumpiness on the radiation drag efficiency.
First, if $\bar{N}_{\rm int} < 1$,
the drag efficiency is saturated when $\bar{\tau} > 1$.
Since the ISM is highly clumpy in this case, the radiation
from distant sources which have different velocities from
the absorbing clouds can contribute to the radiation drag 
even if $\tau_{\rm T}\gg 1$. 
However, the covering factor of clouds is smaller than unity, and therefore
a large fraction of photons escape from the bulge without intersecting clouds.
The saturation is attributed to these two effects. 
Secondly, if $\bar{N}_{\rm int} \approx O(1)$, 
$M_{\rm BH}/M_{\rm bulge}$ grows with $\tau_{\rm T}$ and
maintains a high level in an optically thick ISM.
Obviously, the behavior in the thick limit is different from the uniform case,
although the situation is the same as the uniform case 
in that almost all of photons are consumed inside the bulge. 
This is again because the clumpiness increases the mean free paths of photons, 
so that the radiation from distant sources enhances the drag efficiency.
Thirdly, when $\bar{N}_{\rm int}$ is much larger than unity,
the distribution of ISM is closer to a uniform distributions and therefore
the difference between the velocity of a star and the velocity of 
an absorbing cloud is closer to zero, so that the drag efficiency 
falls as the uniform case.
To conclude, the combined effects of the high clumpiness of the ISM and
the large covering factor of clouds are necessary for high drag efficiency.

\begin{figure}
\rotatebox{270}{\includegraphics[width=3cm,height=8.0cm,keepaspectratio,clip]{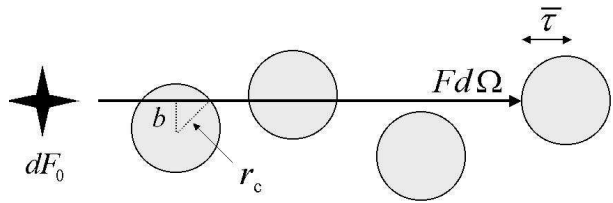}}
\caption[fig2.eps]
{A schematic illustration of the radiative transfer calculations in clumpy ISM.
$b$ and $r_{\rm c}$ are respectively the impact parameter 
and the radius of the clumpy gas clouds. $\bar{\tau}$ is the optical depth
of a cloud.
}
\end{figure}
\begin{figure}
\includegraphics[width=8cm,keepaspectratio]{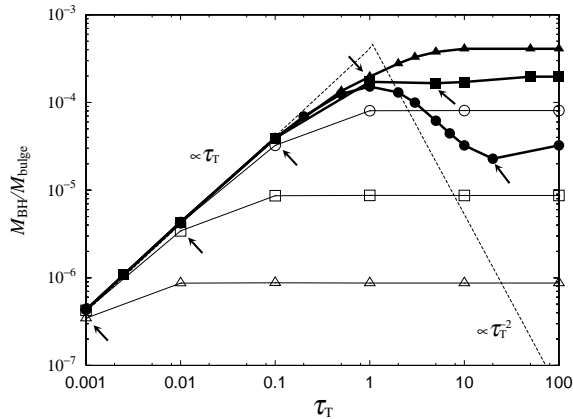}
\caption[fig3.eps]
{The BH-to-bulge mass ratio ($M_{\rm BH}/M_{\rm bulge}$) 
against the total optical depth ($\tau_{\rm T}$) of the bulge.
The thick lines show the results for $\bar{N}_{\rm int}\geq 1$
and the thin lines are those for $\bar{N}_{\rm int}< 1$.
Filled circles denote $\bar{N}_{\rm int}=20$,
filled squares $\bar{N}_{\rm int}=5$, 
filled triangles $\bar{N}_{\rm int}=1$, 
open circles $\bar{N}_{\rm int}=0.1$, 
open squares $\bar{N}_{\rm int}=0.01$, and
open triangles $\bar{N}_{\rm int}=0.001$.
The dashed line is the analytic solution for a uniform ISM 
corresponding to $\bar{N}_{\rm int}\to \infty$, where 
$M_{\rm BH}/M_{\rm bulge} \propto \tau_{\rm T}$ in an optically thin regime
and $M_{\rm BH}/M_{\rm bulge} \propto \tau_{\rm T}^{-2}$ 
in an optically thick regime.
The arrows show the points where the optical depth of a cloud
($\bar{\tau}$) is unity.
For $\bar{N}_{\rm int} \approx 1$, 
the radiation drag efficiency is maximal if $\bar{\tau} \geq 1$.
For $\bar{N}_{\rm int} < 1$, the radiation drag efficiency is 
saturated when $\bar{\tau} \geq 1$.
}
\end{figure}

\section{Black hole-to-bulge mass ratio}
\label{DIS}

The linear relation between the BH mass and the bulge mass
is a direct consequence of the present radiation hydrodynamical mechanism.
The possible mass accreted by the radiation drag is given by 
$M_{\rm max}=\int_{0}^{t(m_l)}L_{\rm bulge}(t)/c^{2}dt$.
Incorporating stellar evolution based on the present single starburst
model, we find $M_{\rm max}=1.2\times 10^{-3} M_{\rm bulge}$.
As shown above, the drag efficiency is 
sensitively dependent on the density distributions of the ISM.
In a highly clumpy ISM where $\bar{N}_{\rm int}$ is smaller than a few, 
the BH-to-bulge mass ratio $M_{\rm BH}/M_{\rm bulge}$ 
can be expressed by
\begin{eqnarray}
\frac{M_{\rm BH}}{M_{\rm bulge}}&=&\eta \frac{M_{\rm max}}
{M_{\rm bulge}}(1-{\rm e}^{-\tau_{T}}) \\
{}&=& 1.2\times 10^{-3}\eta (1-{\rm e}^{-\tau_{T}}), 
\end{eqnarray} 
where $\eta$ gives the radiation drag efficiency.
From the above analysis, $\eta$ is found to be maximally 0.34
in the optically thick limit, and then
$M_{\rm BH}/M_{\rm bulge}=4.1 \times 10^{-4}$.

The present model may be too simple in some respects 
to compare the results closely with the observations.
For a more precise prediction of $M_{\rm BH}/M_{\rm bulge}$, 
it seems necessary to consider further
effects which have not been incorporated in this simple model.
In the present analysis, we assumed an initial coeval event of star formation,
and the subsequent passive evolution without further star formation episodes.
The radiation drag efficiency is basically determined by the total number of
photons which are emitted from sources and absorbed by clouds 
during the whole history of the bulge.
The recycling of the ISM for star formation generates more photons and
therefore could enhance the mass ratio 
roughly by a factor of 2 (Umemura 2001).
Also, in realistic situations, the radiation drag 
is not likely to remove completely the angular 
momentum of stripped gas, and also stripped gas may be mixed with
ISM having appreciable angular momentum. 
Although the detailed processes are not very clear at present,
a little leftover angular momentum may lead to the formation of
a viscous accretion disk around a BH, which could ignite QSO activity 
with nearly Eddington luminosity (Umemura 2001).
If QSO activity is triggered, the radiation from the QSO can
also exert the drag force. This effect can enhance the mass ratio
by maximally a factor of 1.7 (Umemura 2001).
Furthermore, if a galactic wind occurs, 
it might cause a transition from
an optically thick starburst phase to an optically thin QSO phase,
and simultaneously reduce the drag efficiency.

If the above effects are taken into account, 
the mass ratio could be maximally
$M_{\rm BH}/M_{\rm bulge}=1.4 \times 10^{-3} $.
This is compared with the recent observational data in Fig. 4,
combined with the result by a single-starburst model, 
$M_{\rm BH}/M_{\rm bulge}=4.1 \times 10^{-4}$.
In Fig. 4, the hatched area is the prediction of the present analysis.
As seen in this figure, the present radiation hydrodynamical model 
can roughly account for the observational data,
although the prediction falls slightly short of the observed median mass ratio.
The prediction is smaller by a factor of six than 
$\langle M_{\rm BH}/M_{\rm bulge}\rangle =0.006$ 
by Magorrian et al. (1998), while it is comparable to
$\langle M_{\rm BH}/M_{\rm bulge}\rangle =0.001$
by Merritt \& Ferrarese (2001a).

Another important effect is the geometrical dilution.
In previous works, the radiation drag efficiency would be strongly subject to 
the effects of geometry
(Umemura, Fukue, \& Mineshige 1997, 1998; Ohsuga et al. 1999).
A large fraction of emitted photons can escape from a disc-like system 
and thus the radiation drag efficiency is considerably reduced.
Recent observation have revealed that 
the BH mass fraction is significantly smaller than 0.001 in disc galaxies 
(Salucci et al. 2000; Sarzi et al. 2001).
Geometrical dilution may be a reason for the observational fact
in disc galaxies, but the quantitative details are not clear
before such an aspherical system is actually simulated.
\begin{figure}
\includegraphics[width=8cm,height=5.5cm]{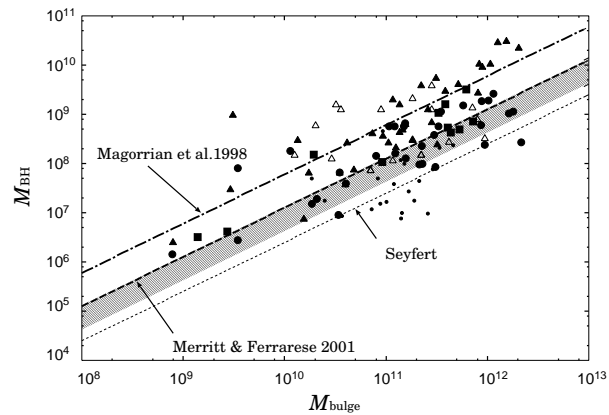}
\caption[fig4.eps]
{The relation between the BH mass and the bulge mass.
The vertical axis is the BH mass, and 
the horizontal axis is the bulge mass, in units of $M_{\odot}$.
Recent observational results are plotted by symbols.
The filled triangles denote elliptical galaxies from Magorrian et al. (1998).
The filled squares show  elliptical galaxies from
Ho (1999), Ferrarese \& Merritt (2000), Kormendy \& Ho (2000), 
and Sarzi et al. (2001).
The filled circles denote elliptical galaxies from Merritt \& Ferrarese (2001a).
The small dots denote Seyfert galaxies from Ho (1999), Wandel (1999), and 
Gebhardt et al. (2000a), and the open triangles show QSOs from Laor (1998).
The relation of Magorrian et al. (1998) is $M_{\rm BH}=0.006M_{\rm bulge}$, 
which is shown by a dot-dashed line;
the relation of Merritt $\&$ Ferrarese (2001a) is $M_{\rm BH}=0.001M_{\rm bulge}$, 
which is shown by a dashed line; and
the relation for Seyfert galaxies is $M_{\rm BH}=2.5\times 10^{-
4}M_{\rm bulge}$ (Sarzi et al. 2001), which is shown by a thin dashed line.
The hatched area shows the prediction of this papaer. 
The lower bound is a single-starburst model 
($M_{\rm BH}=4.1\times 10^{-4}M_{\rm bulge}$). The upper bound is the model 
incoporating the effect of the recycling of star formation and AGN activity
($M_{\rm BH}=1.4\times 10^{-3}M_{\rm bulge}$).
}
\end{figure}

\section{Conclusions}
\label{CON}
By assuming a simple model of a bulge,
we have investigated the mutual effect between the clumpiness
of interstellar medium and the optical depth on 
the radiation drag efficiency for the angular momentum transfer.
In a clumpy interstellar medium, we have accurately solved 3D radiation transfer
to calculate the radiation drag force by the rotating bulge stars. 
We find that the radiation drag efficiency is 
sensitively dependent on the density distribution of the ISM 
in an optically thick regime.
The efficiency drops by a factor of $\tau_{\rm T}^{-2}$ in a uniform ISM, 
while the efficiency turns out to be almost constant 
at a high level in a clumpy ISM.
Also, the radiation drag efficiency falls as the covering factor 
of clouds becomes smaller than unity.
Hence, for the radiation drag to work effectively, 
it is necessary that the covering factor is larger than unity 
and that the distributions of ISM is highly clumpy.
The present radiation hydrodynamical mechanism accounts for 
the linear relation between the mass of a supermassive black hole and 
the mass of a galactic bulge. 
The range of the predicted mass ratio is  
 from $M_{\rm BH}/M_{\rm bulge}=4.1 \times 10^{-4}$ 
to $M_{\rm BH}/M_{\rm bulge}=1.4 \times 10^{-3}$.
Although we have adopted a highly simplified model,
the result roughly accounts for the recent observational trends.
Thus, it seems that more elaborate calculations
deserve to be performed in the present context.
In the future, we should consider further several effects
which have not been incorporated in the present analysis,
and which can influence the mass ratio significantly: e.g. 
the realistic chemical evolution and geometrical dilution.

\section*{Acknowledgments}
We thank M. Mori, T. Nakamoto and H. Susa for helpful discussion.
We are grateful to M. Hayashi and K. Ohsuga for many useful comments.
We also thank the anonymous referee for valuable commments.
Numerical simulations were performed with facilities at Center of Computational Physics, University of Tsukuba.
This work was supported in part by the Grant-in-Aid of the JSPS, 11640225.

\label{lastpage}

\end{document}